\documentclass[]{spie}  
\usepackage[]{graphicx}
\usepackage{textcomp}
\usepackage{array}
\usepackage{color}

\title{Roughness tolerances for Cherenkov telescope mirrors} 

\author{K.~Tayabaly\supit{1,2}, D.~Spiga\supit{1}, R.~Canestrari\supit{1}, G.~Bonnoli\supit{1}, M.~Lavagna\supit{2}, G.~Pareschi\supit{1}
\skiplinehalf
\supit{1} INAF/Brera Astronomical Observatory, Via Bianchi 46, 23807 Merate, Italy\\
\supit{2} Politecnico di Milano, Via La Masa 1, 20156 Milano, Italy}

\authorinfo{contact author: E-mail: \underline{kashmira.tayabaly@brera.inaf.it}, Telephone: +39 039 8943740} 
 
\begin{document} 
\maketitle 

\begin{abstract}
The Cherenkov Telescope Array (CTA) is a forthcoming international ground-based observatory for very high-energy gamma rays. Its goal is to reach sensitivity five to ten times better than existing Cherenkov telescopes such as VERITAS, H.E.S.S. or MAGIC and extend the range of observation to energies down to few tens of GeV and beyond 100 TeV. To achieve this goal, an array of about 100 telescopes is required, meaning a total reflective surface of several thousands of square meters. Thence, the optimal technology used for CTA mirrors' manufacture should be both low-cost ($\sim$1000 euros/m$^2$) and allow high optical performances over the 300-550~nm wavelength range. More exactly, a reflectivity higher than 85$\%$ and a PSF (Point Spread Function) diameter smaller than 1~mrad. Surface roughness can significantly contribute to PSF broadening and limit telescope performances. Fortunately, manufacturing techniques for mirrors are now available to keep the optical scattering well below the geometrically-predictable effect of figure errors. This paper determines first order surface finish tolerances based on a surface microroughness characterization campaign, using Phase Shift Interferometry. That allows us to compute the roughness contribution to Cherenkov telescope PSF. This study is performed for diverse mirror candidates (MAGIC-I and II, ASTRI, MST) varying in manufacture technologies, selected coating materials and taking into account the degradation over time due to environmental hazards. 
\end{abstract}


\section {Introduction}\label{sec:intro}
	
In the last decades Imaging Atmospheric Cherenkov Telescopes (IACT) such as  MAGIC, H.E.S.S. and VERITAS have significantly contributed to investigate and understand the astrophysical sources of Very High Energy (E $\geq$ 30 GeV) gamma rays, such as  supernova remnants, X-ray binaries, micro-quasars, blazars and radiogalaxies. The Cherenkov Telescope Array (CTA) is the next generation of IACTs observatory; under development and construction by a large consortium involving  approximately 30 countries, 150 institutions and more than 1000 scientists, this state-of-the-art facility is expected to have a sensitivity limit 10 times better than the current standard, as well as significant improvements in angular and energy resolution. Obviously, the telescopes design parameters will be crucial to achieve the goals of IACTs. Notably, the total reflective area, determining the amount of light that can be collected, hence establishes the energy threshold of the incoming Gamma photons. 

Split in two sites at $\pm$30~deg of latitude to survey the whole sky, the CTA observatories will consist of three classes of telescopes in the open-air: Large Sized (LST, dish diameter of 23 m), Medium Sized (MST, dish diameter of 12 m) and Small Sized Telescopes (SST, dish diameter of 4 m, of which ASTRI (Astrofisica con Specchi a Tecnologia Replicante Italiana) is a prototype developed by INAF-OAB \cite{ASTRIPrototype}) to observe effectively a large range of energies from a few tens of GeV to beyond 100 TeV. Each site will be equipped with 4 LST, and about 25 MST. The Southern site, more convenient for observation of the Galactic Plane, will also be equipped with approximately 70 SSTs. Therefore, the whole observatory requires producing and integrating several thousands of square meters of mirror panels, chosen as the best trade-off among performance, cost, and durability. The mirror production process should pay particular attention to minimize figure errors, as well as making any scattering from microroughness negligible. In this paper, using a phase shift interferometer, we characterize the micro roughness of several Cherenkov mirrors varying in coating material, manufacturing techniques, and time exposure to real environment and compare their performances. Using the well established first-order scattering theory\cite{Stover}, we predict the PSF of those mirrors in the MST detector plane and therefore compare these various mirrors' surface finish performances. The study is finally pushed further to evaluate the surface finish tolerance for CTA-MST mirrors. 

\section{Design and fabrication of Cherenkov mirrors}\label{sec:designCTA}
	\subsection{Optical design criteria}\label{sec:criteria}

As mentioned in the CTA design concepts \cite{DesignConceptsCTA}, the optical design of the telescope should duly consider several criteria defined below:
	
\begin{description}
	\item[Point Spread Function (PSF):] this indicates the ability of the reflector to concentrate the light emitted by the air showers onto the pixels of the photon detection system that should be of a few arc minutes. Often, the criteria used to characterize the PSF is the W80 diameter, which corresponds to the angular diameter within which $80\%$ of the integrated PSF is concentrated.
	\item[Field of View (FoV):] large FoVs are clearly desirable in order to enable:
	\begin{itemize}
		\item the detection of high-energy showers with large impact distance; 
		\item the efficient study of extended sources and of diffuse emission regions;
		\item large-scale surveys of the sky and the parallel study of many clustered sources.
	\end{itemize} 
	However, the PSF should be maintained as constant as possible throughout the FoV to ensure a good image reconstruction particularly for large impact distant events.
	\item[Local reflectivity:] reflectivity profile is key to enhance the optical system efficiency and to overcome background noise contribution, so guarantee a high signal-to-noise ratio.
	\item[Time dispersion:] the different light paths through the telescope introduce dispersion in the arrival time of photons on the camera. This time delay should not exceed the intrinsic width of about 3 ns of the Cherenkov light pulse from a gamma-ray shower.
\end{description}

\subsection{CTA-Medium Sized Telescope optical design and requirements}\label{sec:requirements}
\begin{figure}[h!]
	\centering
	\includegraphics[width=.6\columnwidth]{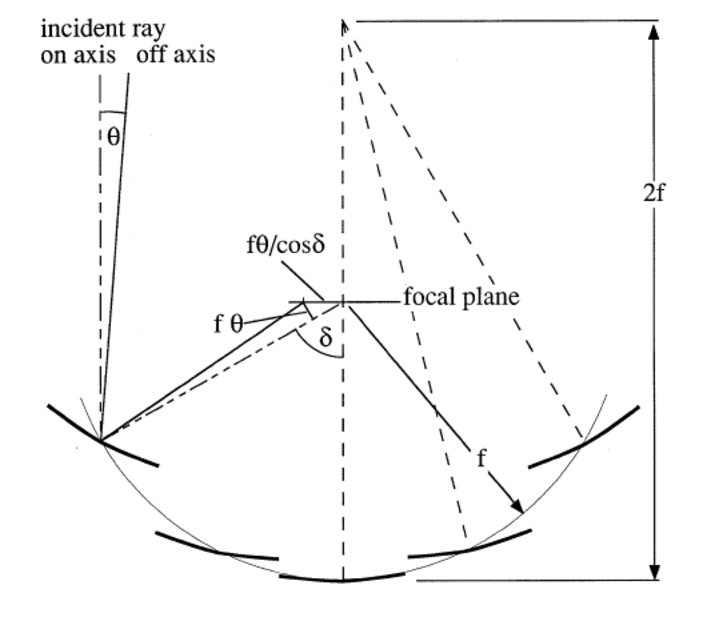}
	\caption{Schematic of the Davies-Cotton design.}
	\label{fig:DC-design}
\end{figure}

The reflector of Cherenkov Telescopes is usually segmented into individual mirrors with a spherical profile. For the optics layout, most current instruments use either a parabolic reflector or a Davies-Cotton design. 
In a parabolic layout, mirrors are arranged on a dish with a paraboloidal profile and the focal length of the mirror facets varies with the distance from the optical axis. In the Davies-Cotton layout, generated originally for solar concentrators \cite{DaviesCotton}, all reflector facets have the same focal length \textit{f}, identical to the focal length of the telescope. The mirror facets are arranged on a dish having a spherical profile of radius \textit{f} (Fig.~\ref{fig:DC-design}). Consequently, the optical surface is discontinuous. Both approaches provide a point-like focus for rays parallel to the optical axis, but suffer from significant aberrations for off-axis rays. While off-axis aberrations are even more accentuated in the parabolic layout, it has the advantage to be isochronous. Comparatively, the Davies-Cotton layout introduces time dispersion to the photons arrival time at the camera, but this does not exceed the intrinsic spread of the Cherenkov wavefront as long as the reflector diameter is smaller than 15~m in diameter. This is why only the parabolic layout option was available for the large MAGIC telescopes aperture (17~m in diameter) in order to limit the large time dispersion that is produced by a Davies-Cotton design.

For the reasons mentioned in Sect.~\ref{sec:criteria} and since a Davies-Cotton geometry provides the best imaging over a large FoV, this configuration was selected for the MST. The main parameters of the MST optical design have been defined by the CTA consortium \cite{CTA_MSTspec} and are listed in Tab.~\ref{tab:CTA_MST}.

\begin{table}[h!]
  \begin{center}
    \begin{tabular}{|c|c|}
 \hline
 Dish diameter =12 m &  \textit{f}/\# $> 1.3$ \\
\hline Optical angular resolution = 0.18~deg & Field of view = 7~deg \\
\hline Average specular reflectivity [300-550~nm] $> 85\%$ & Time dispersion $< 0.8$ ns \\
\hline
    \end{tabular}
  \end{center}
   \caption{\label{tab:CTA_MST} CTA-MST primary mirror specifications stated by the CTA Consortium for the optical design.}
\end{table}
\newpage
In order to comply with the MST specification, the mirrors shall follow specific requirements:
\begin{enumerate}
\item{Each segment shall have a spherical surface profile with 32~m in radius of curvature}
\item{Geometrical errors shall be small enough to guarantee a W80 $<$ 0.06~deg, at the nominal focal plane distance of 16~m. In other words, in the image plane the diameter of the geometrical PSF, W80$_{\mathrm{geo}}$, should not exceed 16.8~mm}.
\item{The scattering contribution shall not degrade the PSF in such a way that the 85\% of the light is still within a 0.12~deg angular diameter, which means that the W85 of the total PSF (geometrical errors+ scattering contribution) shall not exceed $\Phi_{\mathrm{PSF}}$ = 33.9~mm, for rays at normal incidence on the mirror.}
\end{enumerate}

\subsection{Manufacturing techniques for Cherenkov mirrors}\label{sec:manufac}
Different technologies have been adopted so far for the production of the Cherenkov segmented mirrors. The different technologies can be divided into two main groups:
\begin{description}
\item[Aluminized ground-glass mirrors:]
\ They are manufactured with standard technique starting from raw blanks. Ground-glass mirror solution has often been preferred (e.g. HEGRA, CAT, HESS and VERITAS) primarily for its technical maturity, but at the cost of a long production time. Moreover, the ground-glass mirrors are quite heavy translating into increasing cost and complexity for the telescope mechanical structure;
\item[Composite sandwich structure mirrors:] 
\ They are lightweight mirrors consisting of composite sandwich structure. In sandwich construction, membranes (such as sheets of steel, aluminum, glass, or plastic) are bonded to both sides of a core material. This type of construction is characterized by high strength-to-weight. Most of the present and future Cherenkov telescopes make use of composite mirrors. The main ones are manufactured by:
\begin{itemize}
\item direct machining of each individual piece: in this case consisting of an aluminum face sheet curved in autoclave to spherical shape and glued to an aluminum honeycomb inside a thin aluminum box making up a so called ``raw blank''. Each individual raw blank is subsequently polished by a diamond milling fly-cutting process \cite{magicImirror, magicIImirrorAL}. The MAGIC-I telescope and part of the MAGIC-II telescope mirrors have been manufactured in this way.
\item replication process from a mold: the mirror elements have a sandwich-like structure where the reflecting and backing membranes are bonded to both sides of an aluminum honeycomb core. The membranes are thin foils of glass, coated for the reflecting one. The sheets of glass are mechanically bent and conformed to the shape of the master at room temperature by means of vacuum suction. This technique is called glass cold slumping \cite{coldslumping} and it has become the baseline for the manufacturing of the entire set of mirrors for CTA. Part of the MAGIC-II telescope mirrors have been manufactured in this way as well \cite{magicIImirrorGLASSpar}.
\end{itemize}
\end{description}

Mirrors substrate are typically coated with a high-reflective surface composed by a layer of evaporated aluminum plus a thin capping layer of quartz for protection against oxidization. The thickness of the quartz, if suitably chosen, also enhances the peak reflectivity between 300-550 nm. This solution guarantees an average reflectivity profile higher than 85\% over the energy band of interest.

A further technological solution consists in using dielectric coating made of multilayers. By optimizing the number, thickness and type of materials composing the multilayer, the reflectivity profile is tailored to meet the reflectivity requirements. In particular a band-pass filter can be achieved which strongly depletes the reflectivity outside the Cherenkov energy band. This solution should guarantee a higher signal-to-noise ratio and further developments are ongoing. 

All those mirrors have different surface profiles and surface roughness behavior over time, more particularly if we consider them on a microroughness level. Scattering introduced by surface microroughness can significantly damage the performances of optical systems, therefore it is important to predict and understand its contribution to the PSF in order to optimize the manufacturing and testing of the optical components. In the smooth surface regime, the well-established Rayleigh-Rice vector perturbation theory enables a direct correlation between the statistical properties of the surface roughness and the scattering contribution to PSF broadening, as described in the next section. 

\section {Rayleigh-Rice vector perturbation theory}\label{sec:RRtheory}	
For optically smooth, clean and front reflective surfaces, the Rayleigh-Rice vector perturbation theory establishes a direct relationship between the surface roughness and the angular distribution of reflected light on the surface under consideration\cite{Stover}. 

	\begin{enumerate}
	\itemsep1em
	\item\textbf{Assumptions}: 
		\begin{enumerate}
		\item Smooth surfaces
		\\ Illuminated by a ray of wavelength $\lambda$, a surface is commonly considered as smooth if its roughness rms $\sigma$ fulfills the condition: 
		\begin{equation}
		\left(\frac{4\pi \sigma}{\lambda}\right)^2 \ll 1.
		\label{eq:smooth}
		\end{equation} 
		Applying this criterion to our case, the surfaces fall into the smooth surface regime if $\sigma \ll$~24~nm over the full frequency range. 
		\item Clean surfaces
		\\We assume the surface to be deprived of dust particles, so only surface roughness is contributing to scattering. 
		\item Front reflective surfaces
		\\ Only the front surface is scattering light, no contribution from bulk material or back surface are taken into account. 
		\end{enumerate}
		
	\item \textbf{Golden Rule}
The Golden rule establishes a one-to-one relationship between the BRDF (Bidirectional Reflectance Distribution Function) of a surface and its PSD (Power Spectral Density) following the equation:
\begin{equation}
 	\mbox{BRDF}(f_x,f_y) = \frac{16\pi^2}{\lambda^4} \cos\theta_{\mathrm i} \cos\theta_{\mathrm s}\, \mbox{PSD}(f_x,f_y),
 	\label{GoldenRule}
 \end{equation}
where $\lambda$ is the incident light's wavelength, $\theta_{\mathrm i}$ the incident light angle with respect to the normal to the surface, $\theta_{\mathrm s}$ the scattered light angle with respect to the normal of the surface in the incident plane, and $\varphi_{\mathrm s}$ the scattering angle in the azimuthal plane, PSD is the power spectral density of the surface (Eq.~\ref{PSDtheo}) as a function of spatial frequencies in the $x$ and $y$ direction:
\begin{equation}
	 f_x = \frac{\sin\theta_{\mathrm s} \cos\varphi_{\mathrm s}-\sin{\theta_{\mathrm i}}}{\lambda}, \hspace{1cm}
	 f_y = \frac{\sin\theta_{\mathrm s} \sin\varphi_{\mathrm s}}{\lambda}	.
	\label{freqxy}
\end{equation}
	
For a two-dimensional surface $z(x,y)$, defined over a square of side $L$, the two-dimensional PSD of the surface is defined as: 	
\begin {equation}
	\mbox{PSD}_{\mathrm{2D}} (f_x,f_y) = \lim_{L\to \infty} \frac{1}{L^2}  \left|\int_{-\frac{L}{2}}^{\frac{L}{2}} \int_{-\frac{L}{2}}^{\frac{L}{2}} z(x,y) e^{-2 \pi i (f_x x+ f_y y)} \, \mbox{d}x \mbox{d}y \right|^2
	\label{PSDtheo}
\end{equation}	
	
\item \textbf{Power on the detector}:
The power distribution on the detector plane is related to the BRDF (and therefore the PSD, via Eq.~\ref{GoldenRule}) of the reflecting surfaces as: 
	\begin{equation}
 P(x_{\mathrm d},y_{\mathrm d}) = P_{\mathrm r} \cos\theta_{\mathrm s}\, \Omega_{\mathrm s} \, \mbox{BRDF}(f_x,f_y)
	\label{PowerDetector}
	\end{equation}
where $P_{\mathrm r}$ is the totally-reflected intensity off the mirror, $\Omega_{\mathrm s}$ the solid angle subtended by a single pixel on the detector, and 
	\begin{equation}
	x_{\mathrm d} = (\lambda f_x + \sin\theta_{\mathrm i}) d_{\mathrm{MD}}, \hspace{1cm} y_{\mathrm d} =\lambda f_y d_{\mathrm{MD}},
		\label{coord_det}
	\end{equation}
are the $x$ and $y$ coordinates on the detector, having denoted by $d_{\mathrm{MD}}$ the mirror-to-detector distance. 
\end{enumerate}	
	
The above summarized theory provides a powerful tool for predicting the power distribution in the detector plane of an optical system,  based on the surface characterization of its optics, in the smooth-surface regime as shown in Eqs.~\ref{GoldenRule}, \ref{PSDtheo}, and \ref{PowerDetector}. Therefore, accurately characterizing the surface topography of smooth reflective surfaces will not only supply a qualitative estimate of those surfaces, but also allow to appraise the contribution of scattering they introduce to the power distribution in the detector plane.

\section {Surface roughness characterization of Cherenkov mirrors} \label{sec:SurfCharac}
The surface finish of Cherenkov mirrors' surfaces varies depending on manufacturing techniques, coating, and time exposure to the environment on the telescope site. This provides an understanding of the evolution of the surface finish and contribution to scattering over time for mirrors put under environmental conditions.
	
The surface microroughness is characterized in terms of microroughness using the MicroFinish Topographer (MFT) \cite{Parks2011,Tayabaly2013}, a Phase Shift Interferometry microscope. In this section, after describing the MFT and its main features, we compare the surface quality of MAGIC I, MAGIC II, CTA-MST and ASTRI-SST mirrors. All of them are Cherenkov mirrors differing in polishing procedure, coating material and aging (amount of time exposed to the open-air when mounted on the telescope on-site).

\subsection{The MicroFinish Topographer for surface roughness measurement}
\begin{figure}[htb]
  \begin{center}
  \begin{tabular}{p{7cm}p{7cm}}
 	{\includegraphics[height = 5cm,width=6.5cm]{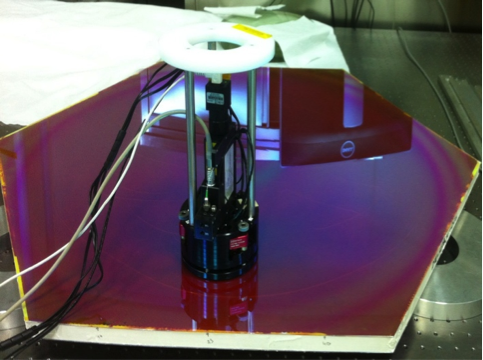}} &{\includegraphics[height = 5cm,width=6.5cm]{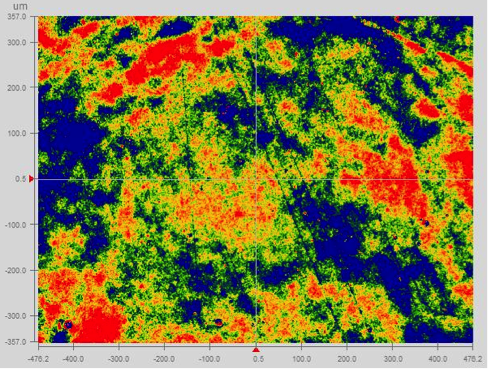}}\\
	\centering \small{A)} & \centering \small{B)}\\
	\end{tabular}
	\caption{A) Picture of the MFT measuring a new ASTRI-Dielectric mirror coated by ZAOT. B) surface topography measured with the MFT, 10$\times$ magnification objective, $\sigma$ = 1~nm.}
	\label{fig: MFTASTRImeas}
	\end{center}
\end{figure}

The MicroFinish Topographer (MFT, Fig.~\ref{fig: MFTASTRImeas},A) is a time shift interferometer equipped with a Nikon 10$\times$ Mirau CF IC Epi Plan DI interferometry objective, a Kohler illumination LED system ($\lambda$= 632.8 nm) and a CCD camera (mod. Flea 2 FL2-08S2M-C) to collect fringes. The area observed by the microscope is 953~$\mu$m $\times$ 714 $\mu$m, with a 0.93~$\mu$m pixel size in the object space. Each surface measurement results from an average of 50 consecutive measurements and the application of a 3$\times$3 smoothing window in order to cancel out environmental vibrations and background noise that is predominant for the highest frequencies observable in this setup. Background noise becomes more daunting as the reflectivity at 632.8~nm is lower -- a likely event with mirrors coated with a narrow bandpass filter -- and the surface is smoother. An example of recorded surface topography is displayed in Fig.~\ref{fig: MFTASTRImeas},B. 

\subsection{Cherenkov mirrors surface topography}
\begin{figure}[htb]
	\centering
	\includegraphics[width=.7\columnwidth]{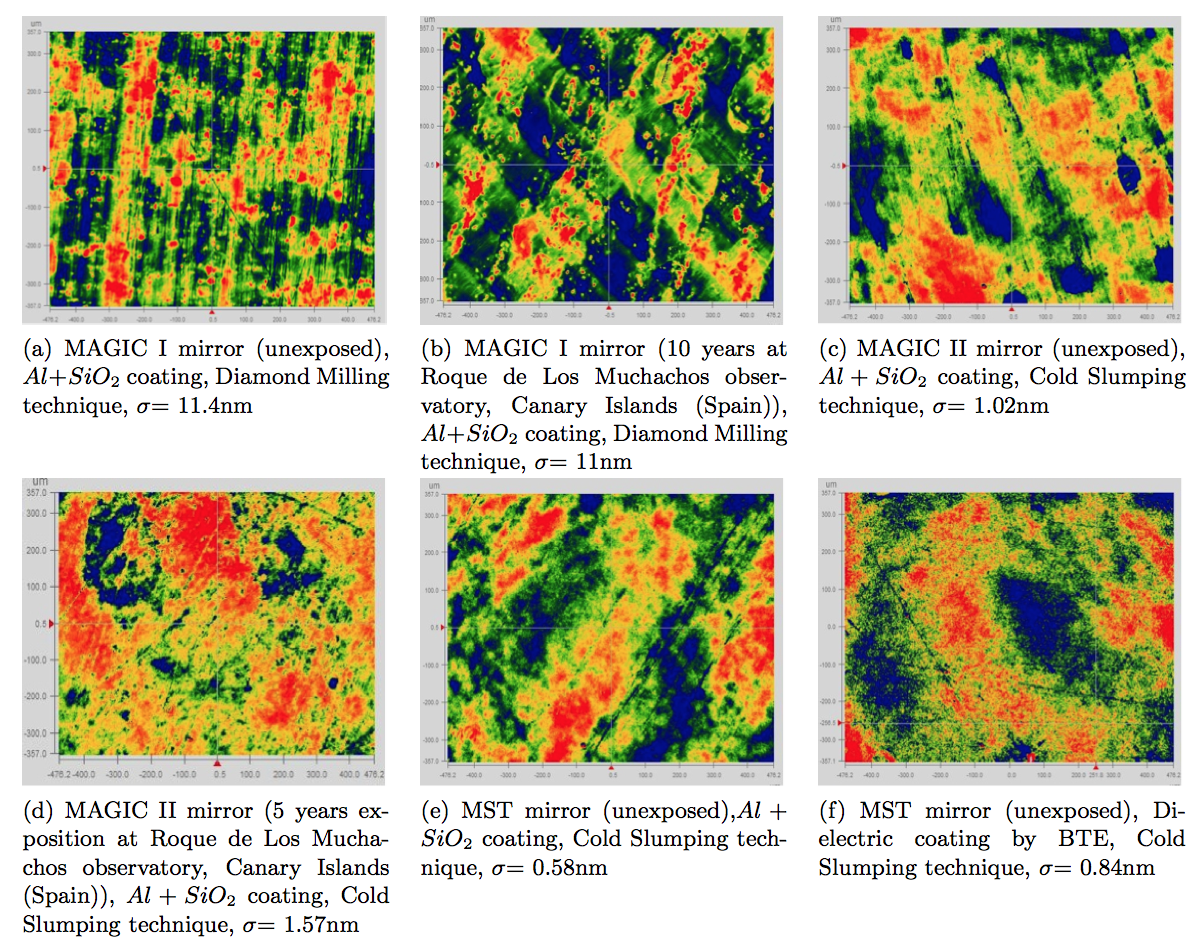}
	\caption{Surface topography of various Cherenkov mirrors measured with the MFT 10$\times$ Mirau objective.}
	\label{fig:SurfaceTopography}
\end{figure}

Using the MFT 10$\times$, the surface topography of diverse types and generations of Cherenkov mirrors were characterized: new and aged mirrors from MAGIC I (aluminum coated mirrors obtained with Diamond Milling technique), new and aged aluminum coated mirrors from MAGIC II, new aluminum coated mirrors from MST, new dielectric coated mirror from MST (designed by BTE GmbH\cite{BTEwebsite}) and ASTRI (designed by ZAOT, Thin Film Coating \cite{ZAOTwebsite}). The mirrors from MAGIC II, MST, and ASTRI were obtained with Cold Slumping technique. The sampled surface topographies are shown in Fig.~\ref{fig:SurfaceTopography}.

A sole comparison of the $\sigma$ value is not sufficient to appreciate the differences observed between new and aged mirrors. In contrast, the surface finish evolution over time caused by the exposure to environmental agents (e.g., rain, salt, moisture, sand, \ldots) becomes more apparent via PSD analysis, as detailed in Sect.~\ref{sec:disc}. 

\subsection{Power Spectral Density comparison}\label{sec:PSDcomp}
\begin{figure}[htb]
	\begin{center}
	\includegraphics[height = 7.2cm]{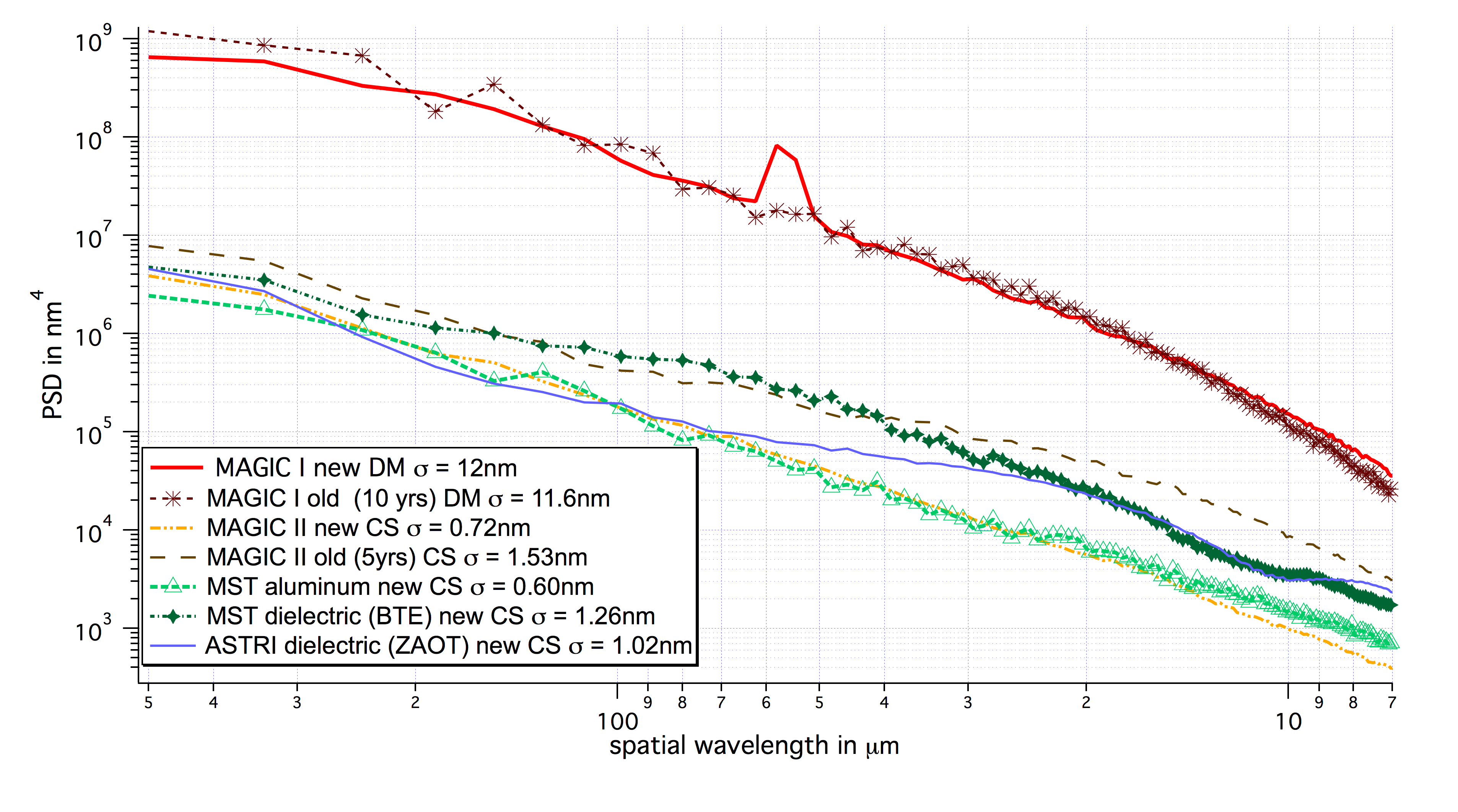}
	\caption{Azimuthally-averaged PSD as a function of spatial wavelength in the range 500-7~$\mu$m for Cherenkov mirrors varying in coating material, manufacture process (DM = Diamond Milling; CS = Cold Slumping ) and time exposure to on-site environment.}
	\label{fig:PSDall}
	\end{center}
\end{figure}

In Fig.~\ref{fig:PSDall} we are comparing the PSD of diverse CTA mirror candidates calculated from the measured surface topography of a wide range of Cherenkov mirrors. With respect to the 2D PSD definition given in Eq.~\ref{PSDtheo}, we have averaged the PSDs over the azimuthal angle $\phi_f$ (where $\tan\phi_f = f_y/f_x$) in order to allow us to compare 1D plots, even this operation cancels out the information related to anisotropy of rough features that are relevant for MAGIC I mirrors (as we see in Sect.~\ref{sec:disc}). The PSDs displayed therein result from the average of the PSD obtained from the measurement of 9 different locations distributed throughout the surface for each of the mirrors considered. 
\newpage


\newpage

\subsection{Discussion}\label{sec:disc}
As anticipated in Sect.~\ref{sec:manufac}, MAGIC I mirrors are aluminum mirrors with a protective layer of quartz produced by Diamond Milling. This technique introduces groove features on the surface that are clearly observable on the surface topography of the MAGIC I mirrors (both new and used, see Fig.~\ref{fig:SurfaceTopography}, A) and B) and Fig.~\ref{fig:PSDall}). The typical spacing of the grooves is in the range 100-50~$\mu$m, with random orientation over the full mirror and peak-to valley amplitude of nearly 30~nm. This structure, already noticed in \cite{CorneliaMS}, is highly contributing to the roughness value of the surface $\sigma_{\mathrm{MI}}$ = 11~nm and the surface scattering properties. In fact, the PSD of MAGIC I mirrors is about 300 times higher than for cold slumped mirrors (Fig.~\ref{fig:PSDall}). 

The MAGIC II mirrors that have been characterized in this paper were manufactured using cold slumping technique. This technique has been improved throughout the years to ensure the production of smoother surfaces. Hence, in the case of aluminum coated mirror, the roughness is improved of more than 50$\%$ between MAGIC II, manufactured in 2008, and MST, manufactured in 2012. They respectively have a roughness of $\sigma_{\mathrm{MII, Al new}}$= 1.0~nm and $\sigma_{\mathrm{MST, Al new}}$  = 0.6~nm, as measured with the MFT. 

The dielectric-coated mirror surfaces have been cold slumped and always exhibit a roughness value about twice (slightly better for ZAOT coating than BTE's, but still quite comparable) as high as cold slumped aluminum mirrors. This could be caused by a well-known phenomenon of surface defect amplification throughout the stack \cite{Spigathesis2005, Canestrari2006}.

Over the spectral range 500-20 $\mu$m, we can notice that MST and new MAGIC II mirrors, cold slumped and coated with aluminum have the same PSD. However, we notice improvement in terms of surface finish for MST at lower spatial wavelengths (up to 15 $\mu$m). As for the dielectric surfaces, ASTRI shows a surface finish as good as aluminum coated mirrors at wavelengths larger than 70 $\mu$m, but it degrades for lower spatial wavelengths where the two dielectric coatings show similar performances. One can also observe the deterioration of the MAGIC II surface after being exposed at the observation site: there is about one decade degradation over almost the full spatial wavelength range considered in terms of PSD amplitude after the mirror was used for 5 years on MAGIC II telescope. As for MAGIC I, given the surface is already quite rough at the production stage, the microroughness degradation is barely distinguishable in the sensitivity window of the MFT. We will see in the next section how those mirrors roughness is expected to contribute to the PSF broadening. 

\section{CTA-MST case study: contribution of scattering to the PSF. Roughness tolerances}\label{sec:EEtolerancing}
	
\subsection{Comparison in terms of encircled energy}\label{subsec: EE_data}	
The Encircled Energy (EE) function, i.e. the PSF integral over the radial coordinate on the detector, is used to describe the focusing performances of an optical system. As such, CTA-MST specifications require that the telescope must focus 85\% of the incoming light within 2/3 of the pixel size, that is, within a diameter of 33.9~mm considering a pixel size of 2" (50.8~mm). The MST design was described in Sect.~\ref{sec:designCTA}. As CTA-MST must focus light with $\lambda$ = 300~nm to 550~nm wavelength range, we compute the expected PSF combining geometrical design and scattering contribution at $\lambda$ = 300~nm, for which scattering introduced by surface roughness is expectedly higher.

If the mirrors had a perfect shape, only the surface roughness would limit the mirror angular resolution. In these conditions, the BDRF would be simply described by Eq.~\ref{PowerDetector}. Assuming for simplicity $\theta_{\mathrm{i}}$ =0 and integrating over the azimuthal coordinate on the detector, the scattering term of the PSF is given by
\begin{equation}
 	\mbox{PSF}_{\mathrm{sc}}(x) = \frac{16\pi^2}{\lambda^3d_{\mathrm{MD}}} \cos^2\theta_{\mathrm s}\, \mbox{PSD}(f_x),
 	\label{eq:PSF}
 \end{equation}
where we have accounted for Eqs.~\ref{freqxy} and~\ref{coord_det}. Moreover, we have denoted with PSD($f_x$) the integral of the PSD (Eq.~\ref{PSDtheo}) over the $\phi_f$ angle (Sect.~\ref{sec:PSDcomp}) in units of nm$^3$. This operation is made possible by the random orientation of roughness features throughout the surface, as observed in Sect.~\ref{sec:disc}, also in the case of MAGIC I mirrors where individual MFT maps exhibit marked grooves. The PSD($f_x$) functions to be used are the ones shown in Fig.~\ref{fig:PSDall}, with an additional factor of $2\pi f_x$.

A real mirror will exhibit also manufacturing errors over larger lateral scales than those typical of roughness, i.e., figure errors that are assumed to be treatable via geometrical optics, returning a geometrical PSF$_{\mathrm{geo}}$. In order to predict the final PSF (or,, equivalently, the EE), we should account for the two species of profile errors using a purely physical optics approach, valid regardless of the spectral range being considered. This is easy to do with grazing incidence mirrors\cite{RaiSpi15}, while for near-normal incidence mirrors a method to include the roughness effect without an excessive increase of the computational complexity\cite{Tayabaly2015} is still under development. In this work, we limit ourselves to add the two contributions of the PSF under a few reasonable hypotheses.

\begin{figure}[h!]
	\centerline{\includegraphics[width=14.5cm]{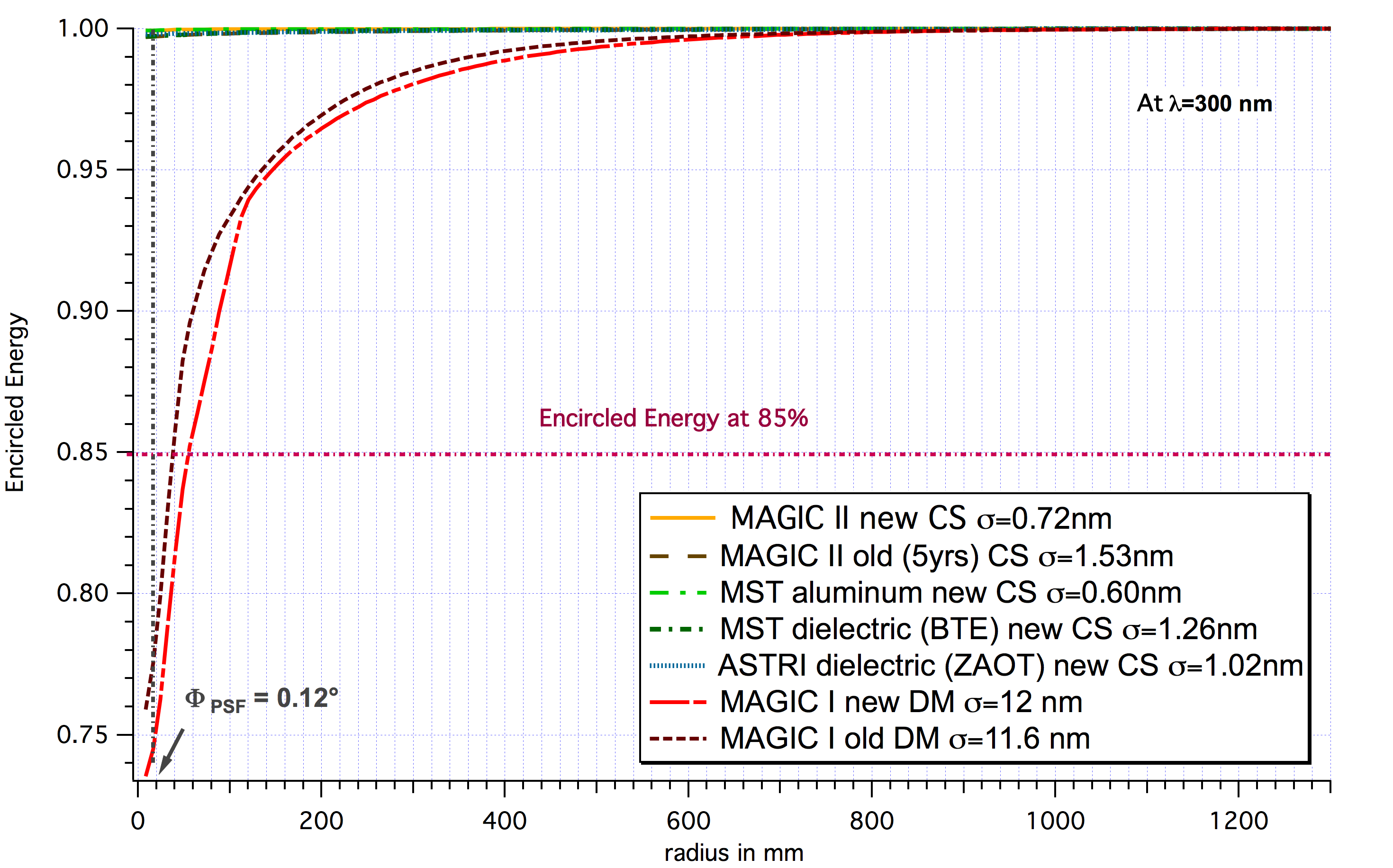}}
	\caption{Encircled Energy for all CTA mirrors considering a monochromatic light at $\lambda$ = 300~nm reaching the mirror surface at normal incidence.}
	\label{fig:EE_CTA_all}
\end{figure}

\begin{figure}[h!]
	\centerline{\includegraphics[width=14.5cm]{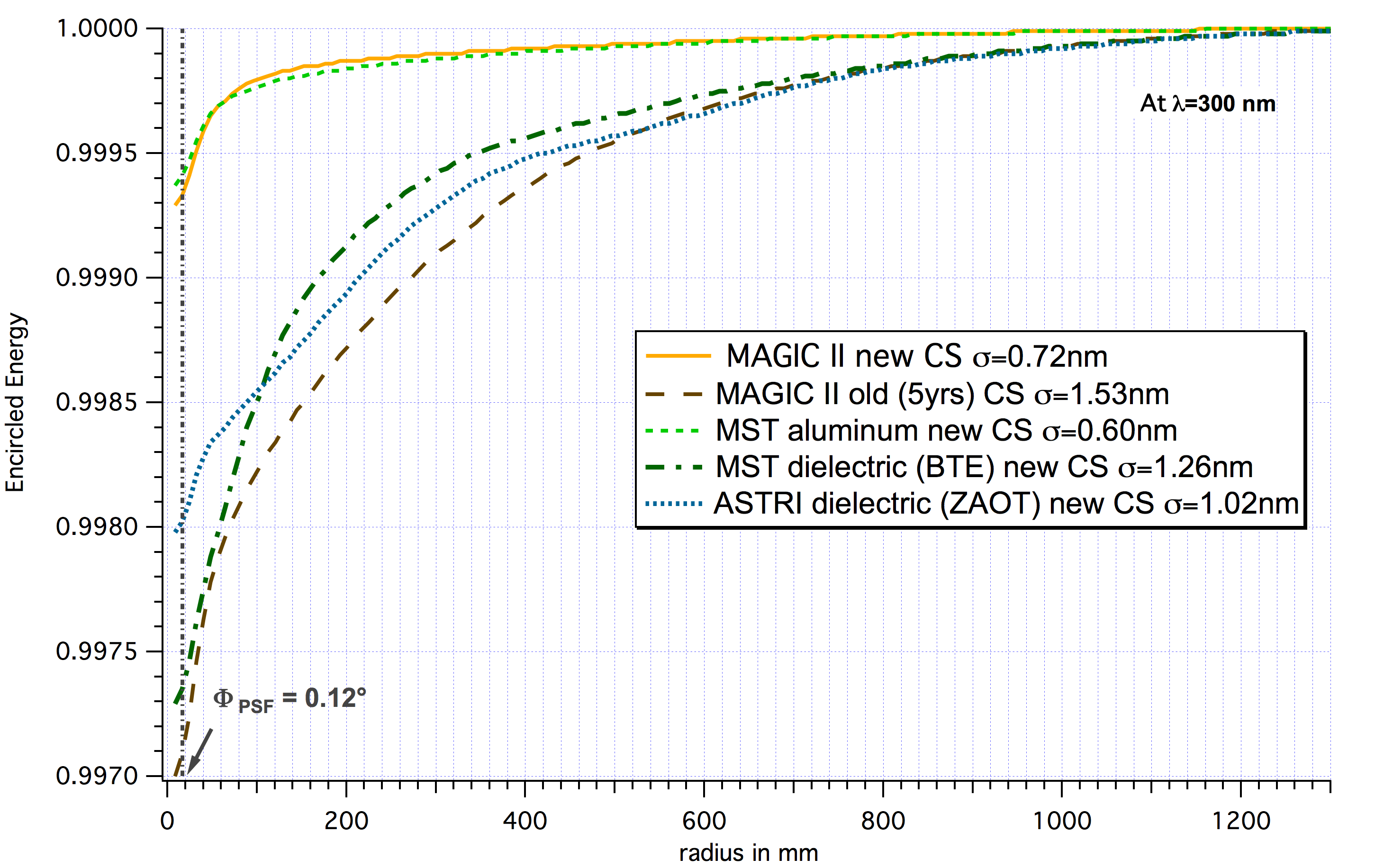}}
	\caption{Encircled Energy for all cold slumped CTA mirrors considering a monochromatic light at $\lambda$ = 300~nm reaching the mirror surface at normal incidence (expanded vertical scale of Fig.~\ref{fig:EE_CTA_all}).}
	\label{fig:EE_CTA_CS}
	\end{figure}

First, we assume that the scattering is caused by a spectral band completely separated from the one identified as geometrical defects, and that the contribution of mid-frequencies to the PSF is negligible. Secondly, we suppose that the roughness entirely fall in the smooth surface limit (Eq.~\ref{eq:smooth}), a condition that is met in all the cases under treatment (Fig.~\ref{fig:PSDall}), so that  Eq.~\ref{eq:PSF} is applicable. This also allows us to consider the scattering as a perturbation of the PSF, i.e., that the PSF broadening imparted by the scattering is much smaller than the one caused by geometric errors:
\begin{equation}
	\mbox{W80}_{\mathrm{sc}} \ll \mbox{W80}_{\mathrm{geo}}. 
 	\label{eq:muchless}
\end{equation}
As discussed in\cite{RaiSpi15} for the case of grazing incidence mirrors the smooth surface approximation, these conditions usually allow us to treat separately the geometric and the scattering regimes, and the total W50 (there named Half-Energy-Width, HEW) is the linear sum of the two contribution. We hereafter assume that the same relation holds for the W80:
\begin{equation}
	\mbox{W80} \approx \mbox{W80}_{\mathrm{sc}} + \mbox{W80}_{\mathrm{geo}},
 	\label{eq:linsum}
\end{equation}
this relation can, however, also be directly derived from the previous assumptions, as shown in Appendix~\ref{sect:app_A}. Considering now the integral of the PSFs predicted from surface roughness measurements of different CTA mirrors meeting the smooth surface requirements, one can compute the expected W80 from Eq.~\ref{eq:PSF}, simply adding linearly a constant term of $\mbox{W80}_{\mathrm{geo}}$ = 16.8~mm (Sect.~\ref{sec:requirements}) and using the computed PSDs (see Fig.~\ref{fig:PSDall}). The resulting EE functions are displayed in Fig.~\ref{fig:EE_CTA_all}, and~\ref{fig:EE_CTA_CS} for the case of the sole cold-slumped mirrors. Finally, in order to fulfill the CTA requirements (Sect.~\ref{sec:requirements}), the total W85, as computed from the EEs, should not exceed the threshold value of $\Phi_{\mathrm{PSF}}$ = 33.9~mm, i.e., 2/3 of the pixel size. This is also equivalent to requiring that no more than 5\% of the scattered power is scattered outside the angular range 0.06~deg $< 2\theta_{\mathrm s}<$ 0.12~deg.
	
The cold slumped mirrors show excellent performances while diamond milled MAGIC I mirrors do not meet CTA requirements as only 75\% of the light is focused within 2/3 of the pixel size. Indeed, Cold Slumped mirrors introduce very little light angular distribution out of the specular beam. For the new generation of CTA mirrors, we can clearly notice an improvement over the different types of technologies used. There is a significant improvement going from diamond milling to cold slumping as we can see in Fig.~\ref{fig:EE_CTA_all}, but this improvement is also noticeable when looking at the different generations of cold slumped mirrors. Indeed, over time, the manufacture process for surface finish has been improved and optimized. Hence, and as for now, the Aluminum coated MST mirrors show the best surface finish. 

However, one should keep in mind that most of the data showed here consider brand-new mirrors while Cherenkov mirrors in use are left under the open air when mounted on the telescope. So, they are subjected to environmental hazards such as salt erosion, sand storms, rain and hail -- to name a few -- that could damage the surfaces. Therefore, one can wonder to which extent these degradations can be tolerated, i.e., how the PSF requirements could  translate in terms of surface microroughness, which is an easy parameter to measure and characterize. In this perspective, in the next section, we translate the PSF requirement stipulated by the CTA Consortium for CTA-MST into a first order roughness tolerance requirement for the mirror panels.

\subsection{Predicting surface degradation to set a roughness tolerance level}
\begin{figure}[htb]
	\centerline{\includegraphics[width=14.5cm]{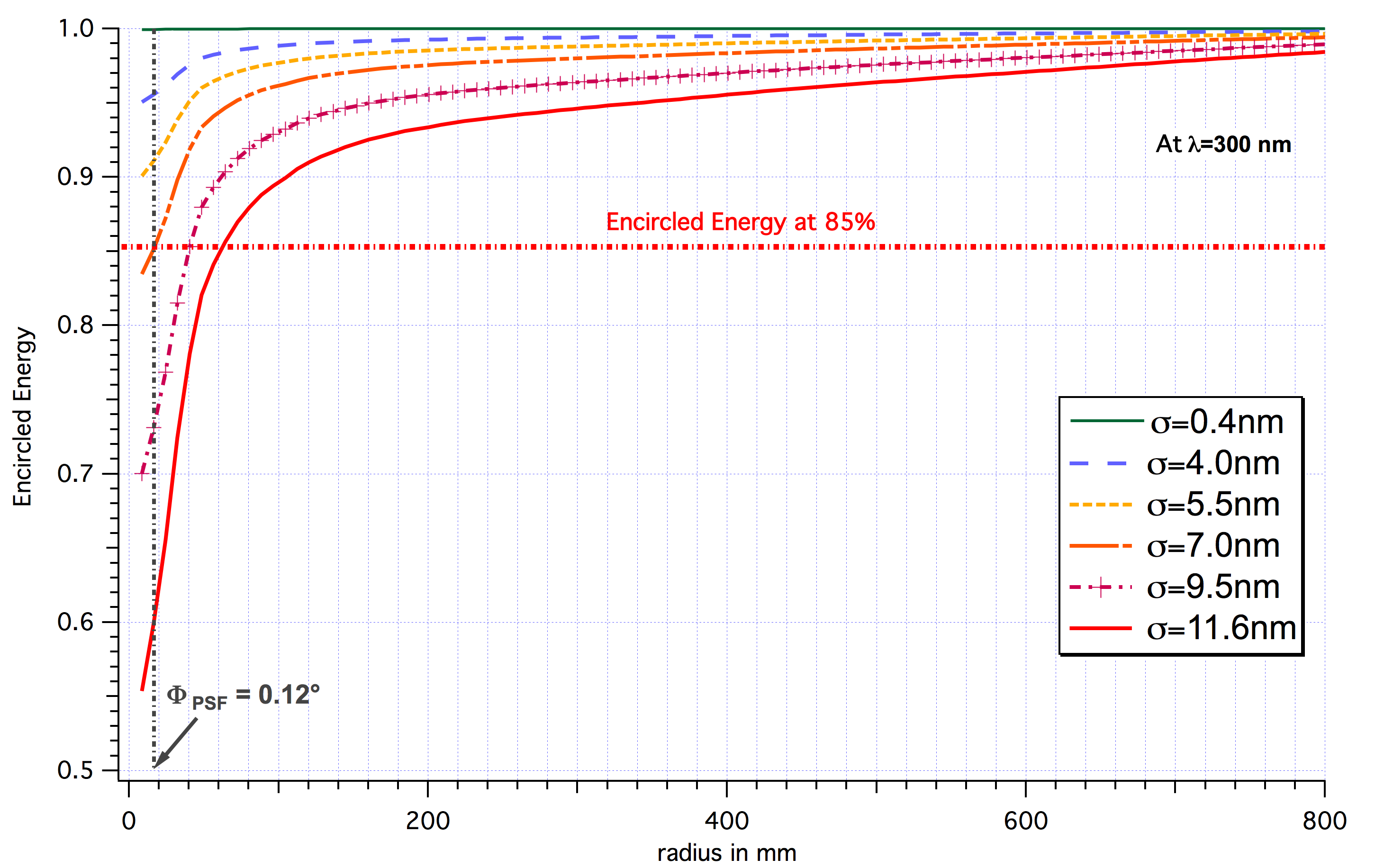}}
	\caption{Simulated Encircled Energy functions, for increasing surface finish degradation ($\lambda$ = 300~nm) at normal incidence. The maximum tolerable rms value is represented by the line at $\sigma$ = 7~nm.}
	\label{fig:EEdeg300nm}
\end{figure}

\begin{figure}[h!]
	\centerline{\includegraphics[width=14.5cm]{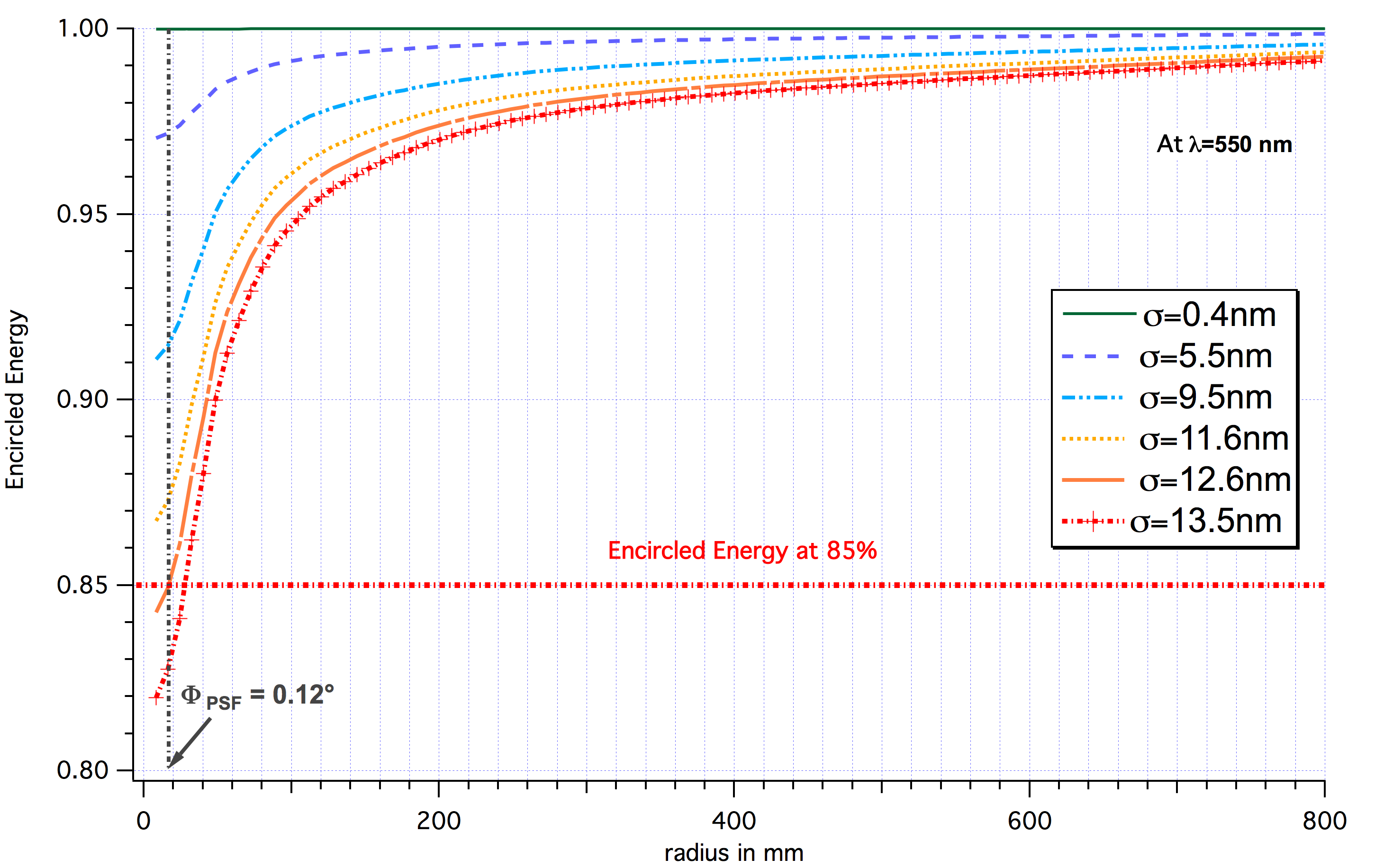}}
	\caption{Simulated Encircled Energy functions, for increasing surface finish degradation ($\lambda$ = 550~nm) at normal incidence.}
	\label{fig:EEdeg550nm}
\end{figure}

Assuming as a baseline the smoothest Cherenkov surface measured, i.e., the new Al+SiO$_{2}$ MST mirror, we simulate in this section a possible PSD degradation to establish a first order tolerance on the mirrors' microroughness. We have magnified the measured roughness by several factors, simulating each time the EE function: the results are shown in Fig.~\ref{fig:EEdeg300nm} at $\lambda$ = 300~nm. Clearly, as the surface roughness degrades, the more difficult it becomes for the mirrors to meet the requirement that W85 $ <\Phi_{\mathrm{PSF}}$. More exactly, {\it the mirrors' surfaces should have a roughness rms $\sigma < $ 7~nm in the spectral wavelength window of 500-7~$\mu$m}. Although not measured yet, assuming a reasonable PSD trend at higher frequencies should not significantly affect the conclusions as the defect amplitudes expectedly become much smaller. Clearly, as the light wavelength increases, the fainter the scattered light and the looser the tolerance roughness becomes (Fig.~\ref{fig:EEdeg550nm}). Nevertheless, the reference value should still be assumed as the one derived from the $\lambda$ = 300~nm, not only as more conservative, but also because as the Cherenkov radiation intensity increases for decreasing $\lambda$.

For a surface equivalent in terms of roughness to MAGIC I mirrors ($\sigma$ = 11.6~nm), this tolerance remains consistent with the  data processing in Sect.~\ref{subsec: EE_data} for MAGIC I mirrors. The roughness tolerance established here is subjected to get tighter if we consider that the scattering introduced by micro roughness adds up to other type of defects or sources of scatters on the mirrors.

\section {Conclusion}
In this paper we have characterized and compared the surface roughness of different Cherenkov mirrors in terms of their PSD, from surface topography data collected with an optical interferometer. Within the smooth surface regime, we could establish prediction in terms of PSF in the particular case of the CTA-MST mirrors and translate the requirements provided by the CTA Consortium in terms of microroughness, which can be easily measured throughout the manufacturing process. As such, to meet the requirement stipulating that W85 shall be less than 0.12~deg at in the light wavelength band 300-550~nm, then $\sigma$ shall be less than 7.0~nm as measured in the 500-7~$\mu$m spatial wavelength range. 

This requirement is met by the cold-slumped mirrors considered as candidates for CTA after manufacturing. However, one should keep in mind that this specification is subjected to get tighter taking into account the environment hazards such as hail, rain, salt erosion or sand storm can damage the surface, as seen with MAGIC I and MAGIC II. Further investigation can and is currently performed to understand more thoroughly the contribution of environmental and meteorological site parameters to surface roughness degradation of Cherenkov mirrors.

\appendix
\section{Derivation of the linear sum of the figure and scattering terms}\label{sect:app_A}
In this appendix we justify the linear sum used to compute the total W80 from geometrical errors and scattering (Eq.~\ref{eq:linsum}). That the two contributions can be added linearly and not, as previously believed, in quadrature is a result already noticed\cite{RaiSpi15} for grazing incidence mirrors under the same assumptions adopted here. One of the results derived there in far-field approximation (a condition widely applicable to the mirror panels being considered in this paper) is that the contributions to the angular resolution degradation from different terms of the profile errors,
\begin{equation}
	z_{\mathrm{err}} = z_{\mathrm{geo}}+z_{\mathrm{sc}}+\cdots,
	\label{eq:zterms}
\end{equation}
add up as a convolution of the respective OTFs (Optical Transfer Functions):
\begin{equation}
	\mbox{OTF} = \mbox{OTF}_{\mathrm{geo}}\otimes \mbox{OTF}_{\mathrm{sc}}\otimes\cdots,
	\label{eq:OTFterms}
\end{equation}
where the $j$-th OTF is the Fourier Transform of the Complex Pupil Function (CPF) of the $j$-th profile error term:
\begin{equation}
	\mbox{CPF}_j = \exp\left[-\frac{4\pi i}{\lambda}z_{j, \mathrm{err}}(\underline{r})\cos\theta_{\mathrm i}\right],
	\label{eq:CPF}
\end{equation}
being $z_{j, \mathrm{err}}$ functions of $\underline{r} = (x_{\mathrm m}, y_{\mathrm m})$, the coordinates on the mirror surface. In near-normal incidence, the same relations between the OTF and the CPF remain valid, with the difference that the definition of the $z_{j, \mathrm{err}}$ error term is slightly different in the sole case of light impinging off-axis\cite{Tayabaly2015}. This is not relevant, however, for the scopes of this work. 

The PSF is the square module of the OTF: if we limit ourselves to the geometric and the scattering term, it can be written as 
\begin{equation}
	\mbox{PSF} = \left(\mbox{OTF}_{\mathrm{geo}} \otimes  \mbox{OTF}_{\mathrm{sc}}\right)\left(\mbox{OTF}_{\mathrm{geo}} \otimes  \mbox{OTF}_{\mathrm{sc}}\right)^*,
	\label{eq:PSF}
\end{equation}
having denoted with * a complex conjugation. Substituting in Eq.~\ref{eq:PSF} the explicit expression of the convolution integrals, it can be developed as
\begin{eqnarray}
	\mbox{PSF} =  \int \mbox{d}^2\underline{h} \int \mbox{d}^2\underline{r}\, e^{-2\pi i \underline{f}\cdot\underline{r}}\,\mbox{CPF}_{\mathrm{geo}}(\underline{r}) \int \mbox{d}^2\underline{r}'\,e^{2\pi i \underline{f}\cdot\underline{r}'} \,\mbox{CPF}^*_{\mathrm{geo}}(\underline{r}')\, \cdot \nonumber\\ 
\cdot \int \mbox{d}^2\underline{r} \,e^{-2\pi i (\underline{f}+\underline{h})\cdot\underline{r}}\,\mbox{CPF}_{\mathrm{sc}}(\underline{r}) \int \mbox{d}^2\underline{r}' \, e^{2\pi i (\underline{f}+\underline{h})\cdot\underline{r}'}\,\mbox{CPF}^*_{\mathrm{sc}}(\underline{r}')  \int \mbox{d}^2\underline{h}' \,e^{-2\pi i (\underline{h'}-\underline{h})\cdot\underline{r}'}
	\label{eq:PSFdev}
\end{eqnarray}
where the integrals in $\underline{r}$ and $\underline{r}'$ are extended to the entire mirror surface, $\underline{f} = (f_x, f_y)$ as defined in Eq.~\ref{coord_det}, and $\underline{h}$ is the frequency lag in the convolution. The product of the two integrals including terms with $\mbox{CPF}_{\mathrm{geo}}$ and $\mbox{CPF}_{\mathrm{sc}}$ are PSF$_{\mathrm{geo}}$ and PSF$_{\mathrm{sc}}$, respectively. We now pose the condition, in general fulfilled in the cases considered in this paper, that the width of the scattering PSF is much smaller than the one of the geometric PSF. In these conditions, the total PSF takes contributions from the integrands only if $|\underline{h}| \ll |\underline{f}|$ at any spatial frequency. In these conditions, also $|(\underline{h'}-\underline{h})\cdot\underline{r}'| \ll 1$ and the last integrand is well approximated by 1. Hence, we remain with
\begin{equation}
	\mbox{PSF} \approx \mbox{PSF}_{\mathrm{geo}} \otimes \mbox{PSF}_{\mathrm{sc}}.
	\label{eq:PSFconv}
\end{equation}

We now want to connect the width of the two PSF terms to the width of the total PSF. We approximate the shape of the PSFs with Lorentzian functions,
\begin{equation}
	\mbox{PSF}(x) = \frac{\Omega}{\pi(x^2+\Omega^2)},
	\label{eq:lorentzian}
\end{equation}
where $\Omega$ is a characteristic width parameter and, to simplify the notation, we have evaluated the spatial frequencies in terms of the distance $x$ from the center of the focal spot on the detector plane. The Lorentzian function is a good model in the characterization of the PSFs of astronomical mirrors, also because the PSD of surface finish defects is in general well approximated by this kind of profile\cite{Stover}. It is easy to see that this PSF is normalized:
\begin{equation}
	\int_{-\infty}^{+\infty}\mbox{PSF}_{\mathrm{sc}}(x) \,\mbox{d}x=\frac{1}{\pi} \left.\arctan \frac{x}{\Omega}\right|_{-\infty}^{+\infty} =1,
	\label{eq:lorentzian_norm}
\end{equation}
and that the related encircled energy function takes at $x = \Omega$ on the value EE$(\Omega) = 1/2$. Hence, W50 = $2\Omega$. Finally, it can be easily proven\cite{Stover} that the Fourier Transform of Eq.~\ref{eq:lorentzian} is 
\begin{equation}
	\mbox{FT(PSF)}(f_x) =\frac{1}{2\pi}e^{-\Omega |f_x|}.
	\label{eq:lorentzian_FT}
\end{equation}
In our case, the two Lorentzian function to be convolved are
\begin{equation}
	\mbox{PSF}_{\mathrm{geo}}(x) = \frac{\Omega_{\mathrm{geo}}}{\pi(x^2+\Omega_{\mathrm{geo}}^2)}, \hspace{1cm} \mbox{PSF}_{\mathrm{sc}}(x) = \frac{\Omega_{\mathrm{sc}}}{\pi(x^2+\Omega_{\mathrm{sc}}^2)}.
	\label{eq:lorentzians}
\end{equation}
The convolution becomes a product in the Fourier Transform space, therefore we have
\begin{equation}
	\mbox{FT}(\mbox{PSF}_{\mathrm{geo}}\otimes\mbox{PSF}_{\mathrm{sc}})(f_x) =\frac{1}{4\pi^2}e^{-(\Omega_{\mathrm{geo}}+\Omega_{\mathrm{sc}}) |f_x|},
	\label{eq:lorentzian_FT}
\end{equation}
and since this is still an exponential function of $|f_x|$, we conclude that the convolution of the two PSFs is still a Lorentzian function with $\Omega = \Omega_{\mathrm{geo}}+\Omega_{\mathrm{sc}}$. Doubling both terms we obtain the final result,
\begin{equation}
	\mbox{W50} = \mbox{W50}_{\mathrm{geo}}+\mbox{W50}_{\mathrm{sc}}
	\label{eq:linear_sum}
\end{equation}
i.e. the diameters including 50\% of the scattered power are to be summed linearly. The same result can be exactly extended to the W80 (or to any width including any other fraction of the total power), simply replacing in Eq.~\ref{eq:lorentzian} the $\Omega$ parameter with
\begin{equation}
	\Omega \rightarrow \frac{\Omega}{\tan\left( \displaystyle0.8\frac{\pi}{2}\right)}
	\label{eq:replace}
\end{equation}
and repeating the passages to obtain the desired result:
\begin{equation}
	\mbox{W80} = \mbox{W80}_{\mathrm{geo}}+\mbox{W80}_{\mathrm{sc}}.
	\label{eq:linear_sum}
\end{equation}

\acknowledgments   
This work is supported by the Italian National Institute of Astrophysics (INAF) and the TECHE and T-REX programs funded by the Ministry of Education, University and Research (MIUR).

\bibliographystyle{spiebib}

\begin{thebibliography}{}

\bibitem{ASTRIPrototype}
Canestrari, R., "The ASTRI SST-2M Prototype: Structure and Mirror," Proc. of ICRC (2013) 

\bibitem{Stover}
Stover, J.C., "Optical Scattering, Measurement and Analysis," SPIE Press (2012)

\bibitem{DesignConceptsCTA}
Actis, M., Agnetta, G., Aharonian, F., et al., "Design concepts for the Cherenkov Telescope Array CTA: an advanced facility for ground.-based high energy gamma-ray astronomy," Exp. Astr., 32, 193 (2011)

\bibitem{DaviesCotton}
Davies, J.M., Cotton, E.S., "Design of the quartermaster solar furnace," Journal of Solar Energy Science Engineering, 1, 16 (1957)

\bibitem{CTA_MSTspec}
The CTA Consortium, "Common Test Facilities and Calibration TDR, vers.3.1", COM/140721 (2015)

\bibitem{magicImirror}
Bigongiari, C., Bastieri,ÊD., Galante,ÊN., Lorenz,ÊE., Mariotti,ÊM., Mirzoyan,ÊR., Moralejo,ÊA., Pepato,ÊA., Peruzzo,ÊL., Saggion,ÊA., Scalzotto,ÊV., Tonello,ÊN., "The MAGIC Telescope reflecting surface," NIM-A, 518, 193 (2004)

\bibitem{magicIImirrorAL}
Doro, M., Bastieri, D., Biland, A., Dazzi, F., Font, L., Garczarczyk, M., Ghigo, M., Giro, E., Goebel, F., Kosyra, R., Lorenz, E., Mariotti, M., Mirzoyan, R., Peruzzo, L., Pareschi, G., Zapatero, J., "The reflective surface of the MAGIC telescope," NIM-A, 595, 200 (2008)

\bibitem{coldslumping}
Canestrari, R., Pareschi, G., Parodi, G., Martelli, F., Missaglia, N., Banham, R.,"Cold shaping of thin glass foils as a method for mirror processing: from basic concepts to mass production of mirrors," Opt. Eng., 52, 2 (2013)

\bibitem{magicIImirrorGLASSpar}
Pareschi, G., Giro, E., Banham, R., Basso, S., Bastieri, D., Canestrari, R., Ceppatelli, G., Citterio, O., Doro, M., Ghigo, M., Marioni, F., Mariotti, M., Salvati, M., Sanvito, F., Vernani, D.,"Glass mirrors by cold slumping to cover 100 m2 of the MAGIC II Cerenkov telescope reflecting surface," Proc. of SPIE, 7018, 70180W (2008)

\bibitem{Parks2011}
Parks, R.E., "MicroFinish Topographer: surface finish metrology for large and small optics," Proc. of SPIE, 8126, 81260D (2011)

\bibitem{Tayabaly2013}
Tayabaly, K., Stover, J.C., Parks, R.E., Dubin, M., Burge, J.H., "Use of the surface PSD and incident angle adjustments to investigate near specular scatter from smooth surfaces," Proc. of SPIE, 8838, 883805 (2013)

\bibitem{BTEwebsite}
www.bte-born.de

\bibitem{ZAOTwebsite}
www.zaot.com

\bibitem{CorneliaMS}
Schultz, C., "Novel All-Aluminum Mirrors of the Magic Telescope Project and Low light level silicon photo-multiplier sensors for future telescopes," Master Degree Thesis at Max Planck Institute (2008)

\bibitem{Spigathesis2005}
Spiga, D., "Development of multilayer-coated mirrors for future X-ray telescopes," Ph.D. dissertation, Universit\'a di Milano-Bicocca (2005) 

\bibitem{Canestrari2006}
Canestrari, R., Spiga, D., Pareschi, G., "Analysis of microroughness evolution in X-ray astronomical multilayer mirrors by surface topography with the MPES program and by X-ray scattering," Proc. of SPIE, 6266, 626613 (2006)

\bibitem{RaiSpi15}
Raimondi, L., Spiga, D., "Mirrors for X-ray telescopes: Fresnel diffraction-based computation of Point Spread Functions from metrology," Astronomy \& Astrophysics, 573, A12 (2015)

\bibitem{Tayabaly2015}
Tayabaly, K., Spiga, D., Sironi, G., Canestrari, R., Lavagna, M., Pareschi, G., "Computation and validation of two-dimensional PSF simulation based on physical optics," Proc. of SPIE, 9577, in press

\end{thebibliography}
{} 

\end{document}